\documentclass[square]{ws-procs975x65}

\def\be{\begin{equation}}
\def\ee{\end{equation}}
\def\bea{\begin{eqnarray}}
\def\eea{\end{eqnarray}}
\def\ba{\begin{array}}
\def\ea{\end{array}}

\begin{document}

\title{Graphene: Relativistic transport in a nearly perfect quantum liquid}

\author{M. M\"uller$^*$}

\address{The Abdus Salam International Center for Theoretical Physics,\\
34014 Trieste, Italy\\
$^*$E-mail: markusm@ictp.it\\
http://users.ictp.it/\~{ }markusm/}

\author{L. Fritz, S.Sachdev}

\address{Department of Physics, Harvard University,\\
Cambridge MA 02138, USA}

\author{J. Schmalian}
\address{Ames Laboratory and Department of Physics and Astronomy, Iowa State
University,\\ Ames, IA 50011, USA}

\begin{abstract}
Electrons and holes in clean, charge-neutral graphene behave like a strongly coupled relativistic liquid. The thermo-electric transport properties of the
interacting Dirac quasiparticles are rather special, being constrained by an emergent Lorentz covariance at hydrodynamic frequency scales. At small carrier density and high temperatures, graphene exhibits signatures of a quantum critical system with an inelastic scattering rate set only by temperature, a conductivity with a nearly universal value, solely due to electron-hole friction, and a very low viscosity. In this regime one finds pronounced deviations from standard Fermi liquid behavior. These results, obtained by Boltzmann transport theory at weak electron-electron coupling, are fully consistent with the predictions of relativistic hydrodynamics. Interestingly, very analogous behavior is found in certain strongly coupled relativistic liquids, which can be analyzed exactly via the AdS-CFT correspondence, and which had helped identifying and establishing the peculiar properties of graphene.
\end{abstract}


\bodymatter

\section{Graphene - a strongly coupled, relativistic electron-hole plasma}
Single layer graphene is a zero-gap semiconductor whose low energy quasiparticles obey the massless Dirac equation~\cite{Semenoff84,Haldane88,Zhou06}. At charge neutrality, the Fermi surface reduces to two inequivalent Fermi points, forming a non-analyticity in the density of states, which can be viewed as a rather simple quantum critical point~\cite{joerg}. On top of that, however, as a consequence of the linear dispersion of the 2d quasiparticles Coulomb interactions are unusually strong. They are only marginally irrelevant under renormalization, flowing only logarithmically to zero with decreasing temperature $T$, see~\cite{guinea}. This is reflected, e.g., in the inelastic scattering rate being proportional to $\alpha^2 T$ ($k_B=\hbar =1$), where $\alpha=e^2/\kappa v_F$ is the (slowly running) dimensionless "fine structure constant" characterizing the strength of Coulomb interactions, where $\kappa$ is the dielectric constant of the adjacent medium and $v_F$ is the Fermi velocity of the linearly dispersing quasiparticles.
This large scattering rate nearly saturates a Heisenberg uncertainty principle for quasiparticles~\cite{subirbook}, according to which the scattering rate is conjectured never to exceed significantly the thermal energy scale. Indeed, upon approaching $\alpha\to O(1)$ one expects a chiral symmetry breaking quantum phase transition towards an insulator~\cite{Laehde} with very different low energy excitations.
Due to the strong marginal interactions, the neutrality point of graphene is very similar to quantum critical points of more complex, strongly coupled materials~\cite{sheehy,qcgraphene,ssqhe,Damle}.
In the quantum critical window, i.e., at small chemical potential of the carriers, $|\mu| < T$, the latter form an interacting "hot" electron-hole plasma with rather unusual transport properties which we discuss below.

\subsection{Anomalously strong Coulomb scattering at charge neutrality}
At finite carrier density, an estimate of the inelastic scattering rate in random phase and Born approximation leads to
\bea
\hbar \tau^{-1}_{\rm inel} \sim {\rm max} (T,|\mu|) \frac{\alpha^2}{(1+\alpha|\mu|/T)^2},
\eea
where $\alpha\approx \alpha(\epsilon)\sim 4/\log(\Lambda/\epsilon)$ denotes the renormalized strength of Coulomb interactions $\alpha$ at a given energy scale $\epsilon={\rm{max}}[\mu,T]$, whereby $\Lambda$ is a UV cutoff.
At finite $\mu$, the scattering rate decreases rather quickly according to the familiar law $T^2/|\mu|$, independent of the interaction strength in the ultraviolet.
The quantum-critical window is clearly distinguished by its strong inelastic scattering rate $\sim \alpha^2 T$, which has several interesting consequences.

As was first pointed out in the context of the superfluid-insulator quantum phase transition~\cite{Damle} the particle-hole symmetric point $\mu=0$ exhibits a finite collision-dominated conductivity, even in the absence of impurities. Indeed, the application of an external electrical field induces counter propagating particle and hole currents, and thus no net momentum. The latter is usually the source of infinite current response unless the momentum decays due to impurities. However, in neutral graphene one finds a disorder-independent conductivity which is solely due to electron-hole friction. Scaling arguments based on the Drude formula, the thermal density of carriers $n_{\rm th}\sim (T/\hbar v_F)^2$, the inelastic scattering rate and a $T$-dependent "effective mass" $m_{\rm eff}\sim T/v_F^2$ suggest a conductivity which grows logarithmically with $T$
\bea
\sigma(\mu=0)\sim \frac{e^2n_{\rm th} \tau_{\rm inel}}{m_{\rm eff}} = \frac{C}{\alpha(T)^2} \frac{e^2}{h}.
\eea
This is indeed confirmed by a microscopic calculation based on the semiclassical Boltzmann equation, which becomes asymptotically exact for $T\ll \Lambda$ where the coupling is $\alpha\ll 1$, yielding the prefactor $C=0.760$~\cite{qcgraphene,kashuba}.

For the same reason as the electrical conductivity remains finite at particle-hole symmetry, the thermal conductivity $\kappa$ diverges at $\mu=0$.
For the case of relativistically invariant systems this has been shown by Vojta {\it et al.}~\cite{Vojta}. $\kappa$ describes the heat current response to a thermal gradient in the absence of an electrical current. Usually, the latter forbids the excitation of a finite, non-decaying momentum, and this ensures a finite heat current response. At particle-hole symmetry, however, the momentum created by a thermal gradient does not carry a net current and is thus not affected by the boundary condition. It follows that within the bulk of a sample a thermal gradient cannot be sustained at $\mu=0$ (see Ref.~\cite{AleinerFoster} for a discussion of $\kappa$ in a sample coupled to leads). For graphene, both relativistic hydrodynamics~\cite{Landau} and Boltzmann theory yield the leading divergence
\bea
\kappa(\mu\to 0)= \frac{\sigma(\mu=0)}{T}\left(\frac{P+\varepsilon}{\rho}\right)^2,
\eea
$P,\varepsilon$ and $\rho$ being the pressure, energy density and charge density of the fluid, respectively. This relation can be interpreted as a relativistic Wiedemann-Franz-like relation between $\sigma$ and $\kappa$.

\subsection{Graphene as a nearly perfect liquid}
A further consequence of the strong Coulomb coupling in graphene, and more generally, of quantum criticality, is the anomalously low value of the shear viscosity $\eta$. Its ratio to the entropy density, $\eta/s$ is the crucial parameter in the Navier-Stokes equation which controls the occurrence of turbulence via the Reynolds number
\bea
\mathrm{Re}=\frac{s/k_{B}}{\eta /\hbar }\times \frac{k_{B}T}{\hbar v/L}%
\times \frac{u_{\mathrm{typ}}}{v_F}\,,
\eea
where $L$ is a typical length and $u_{\mathrm{typ}}$ a typical velocity scale of the electronic current flow. The tendency towards electronic turbulence is stronger the larger is $\mathrm{Re}$. Full-fledged turbulence might require $\mathrm{Re}> 10^4$ in 2d, but interesting, complex flow is already expected at experimentally accessible values $\mathrm{Re}\sim 10^2-10^3$~\cite{viscosity}.

Viscosity having the units of $\hbar\cdot n$ with $n$ a density the ratio has units of $\hbar/k_B$. For massless fermions or bosons, the coefficient of proportionality is essentially the mean free path divided by the thermal de Broglie wavelength. This ratio is usually large, but becomes of order $O(1)$ when the scattering rate tends to saturate Heisenberg's uncertainty relation. For certain strongly coupled relativistic liquids the low value $1/4\pi$ was obtained via the AdS-CFT correspondence. Interestingly, a similarly low value is found for graphene when the weak coupling result
$\eta/s=0.13 \hbar/k_B\alpha^2(T)$ is extrapolated to values of order $\alpha\to O(1)$.~\cite{viscosity}

At charge neutrality, $(\eta/s)^{-1}$ is thus a rather direct measure for the dimensionless interaction strength $g$ (equal to $\alpha$ at $\mu=0$), and can serve as an experimental indicator for the latter. At finite carrier density, $\mu>T$, the ratio $\eta/s$ behaves however as $|\mu/T|^3$, and thus does not directly inform about the strength of the dimensionless coupling $g$. A better measure is instead furnished by the estimate
\bea
\frac{1}{g^2} \propto \frac{k_B \eta}{\hbar \rho}\left(\frac{ T}{\mu}\right)^2.
\eea
One may speculate that in this case $\eta$ is "softly" bounded from below by the restriction $g\lesssim O(1)$.

\subsection{Emergent relativistic hydrodynamics in the infrared}
While the quasiparticles of graphene have relativistic kinematics and the non-interacting theory enjoys Lorentz-invariance, the instantaneous Coulomb interactions obviously break this symmetry.
However, one may expect that at hydrodynamic frequency and length scales, the relativistic invariance re-emerges. This is indeed the case, since the role of interactions mainly consists in establishing local equilibrium, while the relaxation towards global equilibrium is governed by the conservation laws of charge, momentum and energy. The latter essentially remain those of a relativistic fluid, thus fixing the structure of all thermoelectric response functions at small frequencies and wavevectors~\cite{Landau,km,nernst,cyclotron}.

It can be shown~\cite{MHDgraphene} that the specificity of the interactions enters the low-energy hydrodynamic description only via the values of the two dissipative transport coefficients $\sigma$ and $\eta$ discussed above for $\mu=0$. A third coefficient, the bulk viscosity, is very small in graphene due to its near scale invariance.

A microscopic analysis~\cite{MHDgraphene} shows that a hydrodynamic description still applies, even when translational invariance and thus momentum conservation is weakly broken by a small magnetic field and dilute disorder. The necessary condition is that the scattering rates associated with those perturbations are significantly smaller than $\tau_{\rm inel}^{-1}$.
An interesting prediction of the resulting magneto-hydrodynamics is the presence of a collective cyclotron resonance~\cite{nernst,cyclotron}. Its origin is a collective circulating motion of the electron-hole plasma with a frequency proportional to its net charge density $\rho$. In contrast to systems with Galilean invariance with a single type of charge carrier, for which Kohn's theorem ensures a sharp, undamped cyclotron resonance, the cyclotron motion in graphene is damped due to electron-hole friction.

\section{Comparison with strongly coupled liquids solved by AdS-CFT}
Our investigations of interaction effects in graphene were to a large extent stimulated by very similar physical phenomena in strongly coupled critical systems and ultrarelativistic matter such as the quark-gluon plasma.~\cite{AdSCFTviscosity,nernst}
Indeed, it is interesting to confront and compare the properties of graphene with results obtained for relativistic supersymmetric $SU(N)$ gauge theories in $2+1$ dimensions, with which one hopes to catch some of the physically relevant features of the quark gluon plasma. These gauge theories are strongly coupled field theories with an emerging conformal symmetry in the infrared. The AdS$_3$-CFT$_{2+1}$ correspondence stipulates that this theory is exactly dual to an Einstein-Maxwell gravity theory in $3+1$ dimensions, which in the large $N$ limit becomes exactly solvable. Among others, this allows one to compute transport coefficients of the gauge theory~\cite{AdsCFT,AdSCFTviscosity,nernst}
\bea
\sigma = \left(\frac{2}{9}\right)^{1/2} N^{3/2} \frac{e^2}{h}\; ;\quad \quad
s,\eta\propto N^{3/2}T^2\;;\quad \quad
\frac{\eta}{s}=\frac{1}{4\pi}\;.
\eea
These results can be physically interpreted as reflecting properties of $N^{3/2}$ effective degrees of freedom~\cite{Klebanov}, whose strong interactions result in an inelastic scattering time which essentially saturates the Heisenberg-like bound
 $\tau T\geq O(1)$. The AdS-CFT correspondence also confirmed the predictions of relativistic  magnetohydrodynamics, including the cyclotron resonance, as discussed above.~\cite{nernst}

\section{Conclusion}
Graphene exhibits several peculiar transport properties which it shares with other quantum critical systems as well as with ultrarelativistic matter. The crucial ingredients behind these phenomena are anomalously strong Coulomb interactions in 2d and the relativistic kinematics. Experimental verifications at moderately high $T$ should soon be within reach due to the availability of clean suspended graphene.

\end{document}